\documentclass[12pt]{article}
\usepackage{a4wide}
\usepackage[centertags]{amsmath}
\usepackage{amssymb}
\usepackage{amsbsy}
\usepackage{epsfig}
\begin{document}
\begin{flushright}
\begin{tabular}{l}
UWThPh-1996-31
\\
hep-ph/9605248
\\ \\
\end{tabular}
\end{flushright}
\def\thefootnote{\fnsymbol{footnote}}
\begin{center}
\huge
Status of Neutrino Mixing\footnote{
Presented at the
\textit{$7^{\mathrm{th}}$ International Workshop
on Neutrino Telescopes},
Venezia, February 1996.
}
\\
\vspace{0.5cm}
\Large
S.M. Bilenky
\\
\vspace{0.2cm}
\normalsize
Joint Institute for Nuclear Research, Dubna, Russia,\\
and\\
Institut f\"ur Theoretische Physik,
Universit\"at Wien,\\
Boltzmanngasse 5,
A-1090 Vienna, Austria.
\end{center}

\section{Schemes of neutrino mixing}

The investigation of the problem of neutrino
masses and mixing is the central theme of today's neutrino physics.
This problem
is very important for elementary particle physics.
There is a general belief that
its investigation is a way to discover new physics.
The problem of the masses of the
neutrinos has also an exceptional importance for astrophysics.
If the neutrino masses
are in the eV region,
it will allow to solve the problem (or part of the problem)
of dark matter.

After many years of investigations,
the problem of neutrino masses and mixing
is still far from being solved.
At present we have several indications in
favour of nonzero neutrino masses and mixing angle.
These indications come
first of all from the solar neutrino experiments
(Homestake \cite{Homestake},
Kamiokande \cite{Kamiokande},
GALLEX \cite{GALLEX}
and SAGE \cite{SAGE}).
There are also indications
in favour of neutrino mixing
from some experiments on the detection of atmospheric
neutrinos
\cite{Kamiokande-atmospheric,IMB,Soudan}
and from the LSND experiment
\cite{LSND}
in which
beam-stop neutrinos were detected.

In the future with new experiments
Super-Kamiokande \cite{SK},
SNO \cite{SNO},
CHORUS \cite{CHORUS},
NOMAD \cite{NOMAD},
CHOOZ \cite{CHOOZ},
ICARUS \cite{ICARUS},
MINOS \cite{MINOS},
COSMOS \cite{COSMOS}
and many others we
can expect a real progress in the investigation
of the problem of neutrino masses,
neutrino mixing and neutrino nature.

I will start with the formulation of the hypothesis of neutrino mixing.
In accordance with all existing experimental data, 
the interaction of neutrinos with matter
is described by the standard $CC$ and $NC$
Lagrangians
\begin{eqnarray}
&&
\mathcal{L}_{I}^{CC}=
-\frac{g}{2 \sqrt 2}
\,
j_{\alpha}^{CC}
\,
W^{\alpha} + h.c
\\
&&
\mathcal{L}_{I}^{NC}=-\frac{g}{2\cos{\theta _W}}
\,
j_{\alpha}^{NC}
\,
Z^{\alpha}
\end{eqnarray}
Here
\begin{eqnarray}
&&
j_{\alpha}^{CC}
=
2 \sum_{l=e,\mu,\tau} \bar{\nu}_{lL}
\, \gamma_{\alpha} \, l_L
+
\ldots
\\
&&
j_{\alpha}^{NC}
=
\sum_{l=e,\mu,\tau} \bar{\nu}_{lL}
\, \gamma_{\alpha} \, \nu_{lL}
+
\ldots
\end{eqnarray}
are the standard charged and neutral currents
(we are interested here only in the
lepton part of these currents).
Let us notice that the
$CC$ interaction
$\mathcal{L}_{I}^{CC}$
determines the
option of flavour neutrinos:
muon neutrino is the particle that produce muon
in the process
$\nu_{\mu}+N \rightarrow \mu^- + X$,
and so on.  
In the $CC$ and $NC$
interactions flavour neutrinos take part.

From LEP data on the measurement
of the total invisible width of decay of $Z$ it
follows that the number of flavour neutrinos $n_{\nu_{f}}$ 
is equal to three \cite{RPP}:
\begin{equation}
n_{\nu _ f}=2.987 \pm 0.016
\end{equation}

The hypothesis of neutrino mixing
is the assumption that a flavour neutrino field
$\nu_{lL}$ is a
\emph{mixture} of fields of neutrinos with definite mass
\begin{equation}
\nu_{lL}=\sum_{i} U_{li} \nu_{iL}
\;,
\qquad
l=e,\mu,\tau
\end{equation}
Here $\nu_i$ is the field of the neutrino with the mass $m_i$ 
and $U$ is a unitary mixing
matrix.

How
many massive light neutrinos exist in nature?
Let us stress that LEP data
do not give us information about the number of
neutrinos
$\nu_i$ with a small mass $m_i$.
In fact,
assuming that $ m_i \ll m_Z $,
for the total invisible decay width of the $Z$
in the
case of neutrino mixing we have
\cite{BGN90}
\begin{equation}
\Gamma_Z^{inv}
=
\sum_l \sum_{i,k} {|U_{li}|}^2 \, {|U_{lk}|}^2
\,
\Gamma^0(Z\rightarrow\nu\bar\nu)
\end{equation}
where $\Gamma^0(Z\rightarrow \nu\bar\nu)$ is the standard 
decay width of the $Z$ into
a pair of massless neutrino and antineutrino.
Due to the unitarity of the mixing matrix,
independently on the number of
light neutrinos we have
\begin{equation}
\Gamma_Z^{inv}
=
n_{\nu_f} \, \Gamma^0(Z \rightarrow \nu\bar\nu)
\end{equation}

The number of massive neutrinos in different schemes of neutrino mixing is
different and is varied from three to six
(see the reviews \cite{BP78,BP87}).
If the neutrino mass term is a Dirac one
\begin{equation}
\mathcal{L}^D
=
- \sum_{l,l'}
\overline{\nu_{l'R}}
\,
M^D_{l'l}
\,
\nu_{lL} + \mbox{h.c.}
\label{07}
\end{equation}
($M^D$ is a complex non-diagonal matrix),
to three neutrino
flavours correspond three massive neutrinos
$\nu_i$ $(i=1,2,3)$ and
the flavour fields
$\nu_{lL} $ are connected
with the massive fields $\nu _{iL}$  by the relation
\begin{equation}
\nu_{lL} = \sum_{i=1}^3 U_{li} \, \nu_{iL}
\end{equation}
Here $U$ is a unitary
$3\times3$
matrix and
$\nu_i$ is the field of the neutrino with mass 
$m_i$.
In the case of a
Dirac mass term,
the Lagrangian is invariant under the global gauge 
transformations
\begin{equation}
\nu_l \rightarrow e^{i\alpha} \, \nu_l
\;,
\qquad
l \rightarrow e^{i\alpha} \, l
\end{equation}
where $\alpha$ is an arbitrary constant.
The invariance under this transformation
means that the total lepton number 
\begin{equation}
L = L_e + L_\mu + L_\tau
\end{equation}
is conserved and that $\nu_i$ are Dirac neutrinos
($\nu_i$ and $\bar\nu_i$ 
have opposite values
of $L$).
Processes like neutrinoless double-beta decay are
forbidden in this scheme.

Dirac neutrino masses can be generated
by the standard Higgs mechanism together
with the masses of all the other fundamental fermions.
Of course, in this case we have
no explanation of the fact that neutrinos are the lightest fermions in
nature.

In the models beyond the Standard Model,
like GUT models,
the total lepton number
$L$ is not conserved.
The most general neutrino mass term that does not conserve
$L$ is the so called Dirac and Majorana
($D+M$)
mass term 
\begin{equation}
\mathcal{L}^{D+M}
=
\mathcal{L}^M_L
+
\mathcal{L}^D
+
\mathcal{L}^M_R
\end{equation} 
Here
\begin{eqnarray}
&&
\mathcal{L}_L^M
=
-\frac{1}{2}
\sum_{l,l'}
{\overline{(\nu_{l'L})^c}}
\,
M^L_{l'l}
\,
\nu_{lL} + \mbox{h.c.}
\\
&&
\mathcal{L}_R^M
=
-\frac{1}{2}
\sum_{l,l'}
\bar{\nu}_{l'R}
\,
M^R_{l'l}
\,
{(\nu_{lR})}^c + \mbox{h.c.}
\end{eqnarray}
where
${(\nu_{lL})}^c=C\bar{\nu}^T_{lL}$
is the charge conjugated component,  
$M^L$
and $M^R$ are complex
non-diagonal symmetrical $3\times3$ matrices
and
$\mathcal{L}^D$ is given by the expression
(\ref{07}).
The fields with definite masses $m_i$
are in this case \emph{six} Majorana fields $\chi_i$.
This corresponds to the fact that
the left-handed as well as the right-handed components
enter in the mass term and
the lepton number $L$ is not conserved.

The flavour neutrino fields $\nu_{lL}$ are connected with 
$\chi_i$ by
the relation
\begin{equation}
\nu_{lL} = \sum_{i=1}^{6} U_{li} \, \chi_{il}
\label{13}
\end{equation}
where the fields $\chi_i$ satisfy the Majorana condition
\begin{equation}
\chi_i = \chi_i^c = C\bar{\chi}_i^T
\label{14}
\end{equation}

The six Majorana fields $\chi_i$ are connected with
the right-handed
components $\nu_{lR}$ by the relation
\begin{equation}
{(\nu_{lR})}^c=\sum^6_{i=1} U_{\bar{l}i} \chi_{iL}
\label{15}
\end{equation}
The mixing matrix $U$ in
Eqs.(\ref{13}) and (\ref{15})
is a $6\times6$ unitary matrix.

In the framework of the $D+M$ mixing scheme there exist the very 
attractive see-saw mechanism
\cite{seesaw}
of neutrino mass generation
that connects the smallness of the neutrino masses
with the violation of the lepton number at the large
(GUT?) scale.

In the simplest case of one generation,
the $D+M$ mass term has the form
\begin{equation}
\mathcal{L}^{D+M}
=
-\frac{1}{2}
\, m_{L} \, \overline{(\nu_{eL})^c} \, \nu_{eL}
- 
m_D \bar{\nu}_R \nu_L
-
\frac{1}{2}
\, m_R \, {\bar\nu_{eR}} \, (\nu_{eR})^c
+
\mbox{h.c.}
\label{16}
\end{equation}

The masses of the two Majorana particles
(eigenvalues of the mixing matrix)
are
given by
\begin{equation}
m_{1,2}=\frac{1}{2} 
\left | m_R + m_L 
\mp \sqrt{{(m_R -m_L)}^2+4m^2_D} \right |
\end{equation}
Let us assume that
\cite{seesaw} 
\begin{equation}
m_L \simeq 0
\;,
\qquad
m_D \simeq m_F
\;,
\qquad
m_R \simeq M_{GUT} \gg m_F
\end{equation}
where $m_F$ is the mass of the charged lepton or up quark.
This assumption means
that the lepton number is violated
(by the last term of Eq.(\ref{16}))
at a scale
that is much larger than the electroweak scale.
In this case,
for the Majorana masses we have
\begin{equation}
m_1 \simeq \frac{m^2_F}{M_{GUT}}
\;,
\qquad
m_2 \simeq M_{GUT}
\end{equation}

In the general case of three generations,
in the spectrum of the masses of
Majorana particles there are three light masses
(masses of neutrinos)
\begin{equation}
m_i \simeq \frac{{(m_F^i)}^2}{M_i}
\;,
\qquad
i=1,2,3
\;,
\label{20}
\end{equation}
and three very heavy masses $M_i \simeq M_{GUT}$. 
From the see-saw formula (\ref{20}) it follows 
(in agreement with experimental data)
that the neutrino masses are much smaller than
the masses of the other fundamental fermions.

If all the Majorana neutrino masses are small,
due to Eqs.(\ref{13}) and (\ref{15}),
transitions $\nu_l \rightarrow {\bar\nu_{l'L}}$ are possible. 
Here $\bar\nu_{l'L}$ is the state of a sterile neutrino
(quantum 
of the right-handed field $\nu_{l'R})$.
The existence of such transitions
would be a clear signature of new physics.
 
We will discuss now methods 
which allow to reveal the
effects of neutrino masses and mixing.
Some latest data will be also presented.

\section{Tritium $\boldsymbol{\beta}$-spectrum}

The investigation of high energy part of the $\beta$-spectra
is the classical method of the measurement of the "electron
neutrino mass".
In most experiments the decay
\begin{equation}
{^3} H \rightarrow {^3} He + e^{-} + \bar{\nu}_e
\end{equation}
is investigated.
This is a superallowed decay and the electron spectrum
is determined by the phase space.
The spectrum of the decay  of the molecular tritium
\begin{equation}
T_2 \rightarrow (T{}^{3}He)^+ + e^{-} + \bar{\nu}_e
\end{equation}
is given by
\begin{equation}
\frac{dN}{dT} =
C \, p \, E
\sum_{i} W_{i} \left( Q_i - T \right)
\sqrt{ \left( Q_i - T \right)^2 - m^2_\nu }
F(E)
\end{equation}
Here $ E=T + m_e $ and $ p $ is the electron energy and momentum,
$m_\nu$ is the neutrino mass,
$W_i$ is the probability of the transition
to the state $i$ of the molecule
$(T{}^3He)^+$,
$Q_i=Q_0-V_i$,
$V_i$
is the excitation energy of the final
molecule, $Q_0$ is the energy release in
the decay of ${}^3H$.
The results obtained in the latest
experiments are presented in the Table 1.

\begin{table}[b]
\label{T01}
\protect\caption{
The upper bounds for electron neutrino mass obtained in the latest 
${^3}H $ experiments.}
\begin{center}
\begin{tabular}{|l|c|c|}
\hline
Group &
$ m_{\nu} $ (eV) &
$ m^2_{\nu} $ (eV$^2$)
\\
\hline
Tokyo & $<13$ & $ -65 \pm 85 \pm 65 $\\
Zurich & $<11$ & $ -24 \pm 48 \pm 61 $\\
Los Alamos & $<9.3$ & $ -147 \pm 68 \pm 41 $ \\
Livermore & $<7$ & $ -130 \pm 20 \pm 15 $ \\
Mainz & $<7.2$ & $ -39 \pm 34 \pm 15 $ \\
Troitsk & $<4.35$ & $ -4.1 \pm 10.9 $\\
\hline
\end{tabular}
\end{center}
\end{table}

As it is seen from the Table 1,
in modern tritium experiments
no indications in favour of nonzero neutrino masses
were obtained.
However, some anomaly was found in these
experiments: in all experiments the average value of $m^2_{\nu}$
is negative (see the last column of the Table 1).
This means that, instead
of the possible deficit of the events at the end of the spectrum
(that would corresponds to a
positive $m^2_{\nu}$),
some excess of events is observed.
This excess is clearly seen in the spectrum measured in the Troitsk
experiment \cite{troitsk}.
According to my knowledge,
there is no
understanding of the origin of this anomaly at the moment.

Another method to reveal
the effects of neutrino masses is the search for
neutrinoless double $\beta$-decay
($(\beta\beta)_{0\nu}$-decay) 
of some even-even nuclei.

\section{Neutrinoless double $\boldsymbol{\beta}$-decay}

The process
\begin{equation}
(A,Z) \rightarrow (A,Z +2) + e^{-} + e^{-}
\end{equation}
is allowed only if neutrinos are massive Majorana
particles.
The Hamiltonian of the process is given by
\begin{equation}
H_I = \frac{G_F}{\sqrt{2}}
\,
2
\left( \bar{e}_L \gamma_\alpha \nu_{eL} \right)
j^{\alpha} + \mbox{h.c.}
\end{equation}
where $ j^{\alpha} $ is the hadron charged current and $ \nu_{eL} $
is a mixed field.
The  ${(\beta \beta)}_{0\nu}$-decay is a process of the second
order of the perturbation theory and
neutrino mixing enters into the matrix
element of the process through the propagator
\begin{eqnarray}
\overset{\rule{0.5pt}{10pt}}{\vphantom{\nu^{T}_{eL}}\nu_{eL}}(x_1)
\hskip-2.4em
\overline{
\vphantom{\overset{|}{\vphantom{\nu^{T}_{eL}}\nu_{eL}}}
\hskip+2.4em\hskip+0.66em
\vphantom{\overset{|}{\nu^{T}_{eL}}}
}
\hskip-0.66em
\overset{\rule{0.5pt}{10pt}}{\nu^{T}_{eL}}
(x_2)
& = & \sum_{i} U_{ei}^2  
\frac{1- {\gamma}_5}{2}
\,
\overset{\rule{0.5pt}{10pt}}{\vphantom{\bar{\chi}_i}\chi_i}(x_1)
\hskip-2.2em
\overline{
\vphantom{\overset{|}{\vphantom{\bar{\chi}_i}\chi_i}}
\hskip+2.2em\hskip+0.5em
\vphantom{\overset{|}{\bar{\chi}_i}}
}
\hskip-0.5em
\overset{\rule{0.5pt}{10pt}}{\bar{\chi}_i}
(x_2)
\,
\frac{1-{\gamma}_5}{2} C
\nonumber\\
& = &-\left(\sum_i U^2_{ei}m_i\right)
\frac{i}{{(2\pi)}^4}
\int d^4 p \,
\frac{e^{-ip(x_1-x_2)}}{p^2-m^2_i}
\, \frac{1-\gamma_5}{2} \, C
\label{26}
\end{eqnarray}
where we have taken into account
that for a Majorana field
\begin{equation}
\chi_i^T = - \bar\chi_i \, C
\end{equation}
For small values of the neutrino masses
we can neglect $m_i^2$ in the denominator of
Eq.(\ref{26}).
Thus, the matrix element of
$(\beta\beta)_{0\nu}$-decay is proportional to
\begin{equation}
\langle m \rangle = \sum_{i} U^2_{ei} \, m_i
\end{equation}
The matrix element $ U_{ei} $ is in general complex.
This means that
some cancellation 
of the contributions of different Majorana
neutrinos to $ \langle m \rangle $ is possible.
To illustrate this statement,
let us
assume that $CP$ is conserved in the lepton sector.
In this case, 
the mixing matrix satisfies the
condition $ U_{ei} = U_{ei}^{*} \eta_i $,
where $ \eta_i = \pm i $ is the
$CP$ parity of the Majorana neutrino with mass $m_i$.
In this case,
for the quantity $ \langle m \rangle $ we have
\cite{mbb}
\begin{equation}
\langle m \rangle
=
\sum_{i} |U_ {ei}|^2 \, m_i \, \eta_{i}
\label{29}
\end{equation}
If the $CP$ parities of the different
$\nu_i$ are different,
there are cancellations in Eq.(\ref{29}).

There are about 40 experiments on the search for neutrinoless double
$\beta$-decay going on.
No indication in favour of this process have
been found so far.
The most stringent limits were obtained in
the experiments
searching for $(\beta\beta)_{0\nu}$ decay of
$^{76}Ge$ and $^{136}Xe$. 
In the
experiment of the Heidelberg-Moscow collaboration
the following lower bound
for the time of life of $^{76}Ge$ was found
\cite{H-M}:
\begin{equation}
T^{0\nu}_{1/2} > 5.1 \times 10^{24} \, \mbox{y}
\end{equation}
From this bound it follows that
$ |\langle m \rangle| < (0.7-2.3) $ eV.
In two
years this collaboration plans to reach the limit
$ T^{0\nu}_{1/2} \simeq
{2\times{10^{25}}} $ y.
The Heidelberg-Moscow collaboration obtained for the
time of life of the allowed decay
$ ^{76} Ge \rightarrow ^{76} Se + e^- + e^- + 
\bar{\nu}_e +\bar{\nu}_e $
the following impressive result:
\begin{equation}
T^{2\nu}_{1/2} =
(1.43 \pm 0.03 \pm 0.13 ) \times{10^{21}} \, \mbox{y}
\end{equation}

For the time of life of the
$(\beta\beta)_{0\nu}$-decay of $^{136}Xe$,
the Caltech-Neuchatel-PSI collaboration \cite{C-N-P}
obtained the following lower
bound:
\begin{equation}
T^{0\nu}_{1/2} > 3.4 \times{10^{23}} \, \mbox{y}    
\end{equation}

In the nearest future,
with many new experiments now in preparation
(NEMO \cite{NEMO} and others),
the sensitivity for the quantity $\langle m \rangle$ is planned to be
approximately one order of magnitude better than the present sensitivity
\cite{MV94}.

I will turn now 
to the discussion of

\section{Neutrino oscillations}

Neutrino oscillations were first considered by
B. Pontecorvo \cite{Pontecorvo}.
If there is neutrino mixing
and the neutrino masses are small
enough,
the state of a flavour neutrino $\nu_l$
with momentum $\vec{p}$ is given by
a \emph{coherent superposition} of the states of massive neutrinos
\begin{equation}
\left| \nu_l \right\rangle =
\sum_{i} \left| i \right\rangle U^*_{li}
\label{33}
\end{equation}
here $\left|i\right\rangle$ is the state of a neutrino with negative
helicity,
momentum $p$ and energy $E_i=\sqrt{m_i^2 +p^2} \simeq
p+m_i^2/2p$.
For the amplitude of the transition
$ \nu_l \rightarrow \nu_{l'}$ during the time $t$,
from Eq.(\ref{33}) we have
\begin{eqnarray}
A_{\nu_l \rightarrow \nu_{l'}} & = &
\sum_{i} \langle \nu_{l'} | i \rangle
e^{-i E_i t} \langle i | \nu _l \rangle =
\nonumber\\ 
& = & \sum_{i} U_{l'i} \, e^{-i E_i t} \, U^*_{li} =
\nonumber\\
& = & e^{-i E_1 t} \sum_{i} U_{l'i} \,
e^{-\frac{i \Delta m^2_{i1}}{2p} R}
\, U^*_{li}
\end{eqnarray}
Here $ \Delta m^2_{i1} = m^2_i - m^2_1 $ and $R \simeq t $
is the distance between the neutrino source and detector.
From this expression
it is obvious that neutrino oscillations can take place if there is neutrino
mixing
($U$ is not a diagonal matrix)
and if at least one neutrino mass
difference squared $\Delta m^2$ satisfies the inequality
\begin{equation}
\Delta m^2 \gtrsim \frac{p}{R}
\end{equation}
with $\Delta m^2$ in $\mbox{eV}^2,R$ in $ m$ and $p$ in MeV.
From this
inequality it follows that
the sensitivity to $\Delta m^2$ of experiments
with neutrinos from different facilities varies from
$ \sim \mbox{eV}^2 $
(accelerators) to $ \sim {10^{-11}} \mbox{eV}^2 $ (the sun).
Thus, as it was first stressed by B. Pontecorvo,
neutrino
oscillations is a very powerful tool
for the investigation of the problem of neutrino masses and mixing.

Many experiments on the search for oscillations
have been done with
neutrinos from reactors and accelerators
(see Ref.\cite{Winter}).
No indication in favour of
neutrino oscillations were found in all experiments,
except the Los Alamos
experiment \cite{LSND}.
In this experiment neutrinos are produced in the decays   
(at rest)
$\pi^+ \rightarrow \mu^+\nu_\mu$,
$\mu^+ \rightarrow e^+ \nu_e \bar{\nu_\mu}$.
There are no $\bar{\nu_e}$'s in the initial
beam
(the estimated relative yield of $\bar\nu_e$ from background
is
$ \simeq 4 \times{10^{-4}}$ at
$ E \geq 36 \, \mbox{MeV} $).
In the LSND experiment 9 events 
$ \bar \nu_e + p \rightarrow e^+ + n $
($ n + p \rightarrow d + \gamma $)
were observed
with a neutrino energy in the interval
$ 36 \, \mbox{MeV} \leq E \leq 60 \, \mbox{MeV} $.
The expected
background is
$ 2.1 \pm 0.9 $ events.
These data were interpreted \cite{LSND}
as an indication in favour of
$ \bar\nu_\mu \rightarrow \bar\nu_e $
oscillations.

The most stringent limits on the parameters that characterize
$\nu_\mu \leftrightarrows \nu_e$
oscillations were obtained in the BNL E776
\cite{BNLE776}
and KARMEN \cite{KARMEN}
experiments.

We have considered \cite{BBGK96}
the mixing of three massive 
neutrino fields and
we have assumed that neutrino masses satisfy hierarchy
\begin{equation}
m_1 \ll m_2 \ll m_3
\end{equation}
with $ \Delta m^2_{21} = m^2_2 - m^2_1 $,
that is relevant for the
suppression of solar $\nu_e$'s.
In this case,
for the probability of the transitions
$\nu_l \rightarrow \nu_{l'}$
($ l' \neq l $)
of terrestrial neutrinos we have
\begin{equation}
P_{\nu_l \rightarrow \nu_{l'}}
=
\frac{1}{2}
\,
A_{\nu_l ; \nu_{l'}}
\left(1-\cos{\frac{\Delta m^2 R}{2 p}}\right)
\end{equation}
where $\Delta m^2=m^2_3 -m^2_1$ is the largest mass difference
squared and the amplitude of oscillations is given by
\begin{equation}
A_{\nu_l ; \nu_{l'}} = 4 \, |U_{l'3}|^2 \, |U_{l3}|^2
\end{equation}
The probability of $\nu_l$ to survive is given by
\begin{equation}
P_{\nu_l \rightarrow \nu_l} = 
1 - \sum _{l'\neq l} P_{\nu_l \rightarrow \nu_{l'}} =
1 - B_{\nu_l ;\nu_l}
\left(1- \cos{\frac{\Delta m^2 R}{2 p}}\right)
\label{38}
\end{equation}
where
\begin{equation}
B_{\nu_l ;\nu_l} = \sum_{l' \neq l} A_{\nu_{l'} ; \nu_l}
= 4 \, |U_{l3}|^2 \left(1 - |U_{l3}|^2\right)
\end{equation}
From this relation we find
\begin{equation}
|U_{l3}|^2 = \frac{1}{2}
\left(1 \pm \sqrt{1 - B_{\nu_l ;\nu_l}}\right)
\label{44}
\end{equation}
At any fixed value of $\Delta m^2 $,
from exclusion plots obtained
in reactor and accelerator disappearance experiments we have
\begin{equation} 
B_{\nu_l ; \nu_l} \leq B^0 _{\nu_l ; \nu_l}
\qquad
(l=e,\mu).
\label{45}
\end{equation}
From Eqs.(\ref{44}) and (\ref{45})
it follows that
\begin{equation}
|U_{l3}|^2 \leq a^0_l
\qquad \mbox{or} \qquad
|U_{l3}|^2 \geq 1 - a^0_l
\end{equation}
where
\begin{equation}
a^0_l = \frac{1}{2}
\left( 1 - \sqrt{ 1 - B_{\nu_l ;\nu_l}^{0} } \right)
\end{equation}

We have used the results of the
Bugey \cite{Bugey95}, CDHS \cite{CDHS84} and CCFR84 \cite{CCFR84}
experiments
and we have considered
the interval
$ 10^{-1} \, \mathrm{eV}^2
\le \Delta m^2 \le
10^{3} \, \mathrm{eV}^2 $.
From these results it follows that the parameters $a^0_e$ and
$a^0_\mu$ are small
($ a^0_e \leq 2 \times 10^{-2} $
in the whole region of 
$\Delta m^2$
and
$ a^0_\mu \leq 10^{-1} $ for
$\Delta m^2 \gtrsim 5\times 10^{-1} \, \mbox{eV}^2 $). 

From solar neutrino data
it follows that the value of $|U_{e3}|^2$
cannot be large.
In fact, for the case of a hierarchy
of neutrino masses
that we have assumed,
the probability of the solar neutrinos to
survive is given by
\cite{SS92}
\begin{equation}
P_{\nu_e \rightarrow \nu_e} = |U_{e3}|^4 +
\left(1 - 
|U_{e3}|^2 \right)^2
P^{1,2}_{\nu_e \rightarrow \nu_e}
\end{equation}
where $P^{1,2}_{\nu_e \rightarrow \nu_e}$
is the survival probability
due to the coupling of
$\nu_e $ with $\nu_1$ and $\nu_2$.
If the parameter
$|U_{e3}|^2$ satisfies the inequality
$ |U_{e3}|^2 \geq 1 - a^0_e $,
from Bugey
data it follows that
$ P_{\nu_e \rightarrow \nu_e} \geq 0.92 $
for all solar
neutrino energies.
Such large value of the survival probability is not
compatible with solar neutrino data.
Thus, we have only two allowed regions
of the values of $|U_{e3}|^2$ and $|U_{\mu 3}|^2$:

\def\theenumi{\Roman{enumi}}

\begin{enumerate}

\item \label{R1}
$ |U_{e3}|^2 \leq a^0_e \ll 1 $
and
$ |U_{\mu3}|^2 \leq a^0_\mu \ll 1 $;

\item \label{R2}
$ |U_{e3}|^2 \leq a^0_e \ll 1$
and
$ 1 \geq |U_{\mu3}|^2 \geq 1 - a^0_ \mu $.

\end{enumerate}

\begin{figure}[tbhp]
\begin{center}
\mbox{\epsfig{file=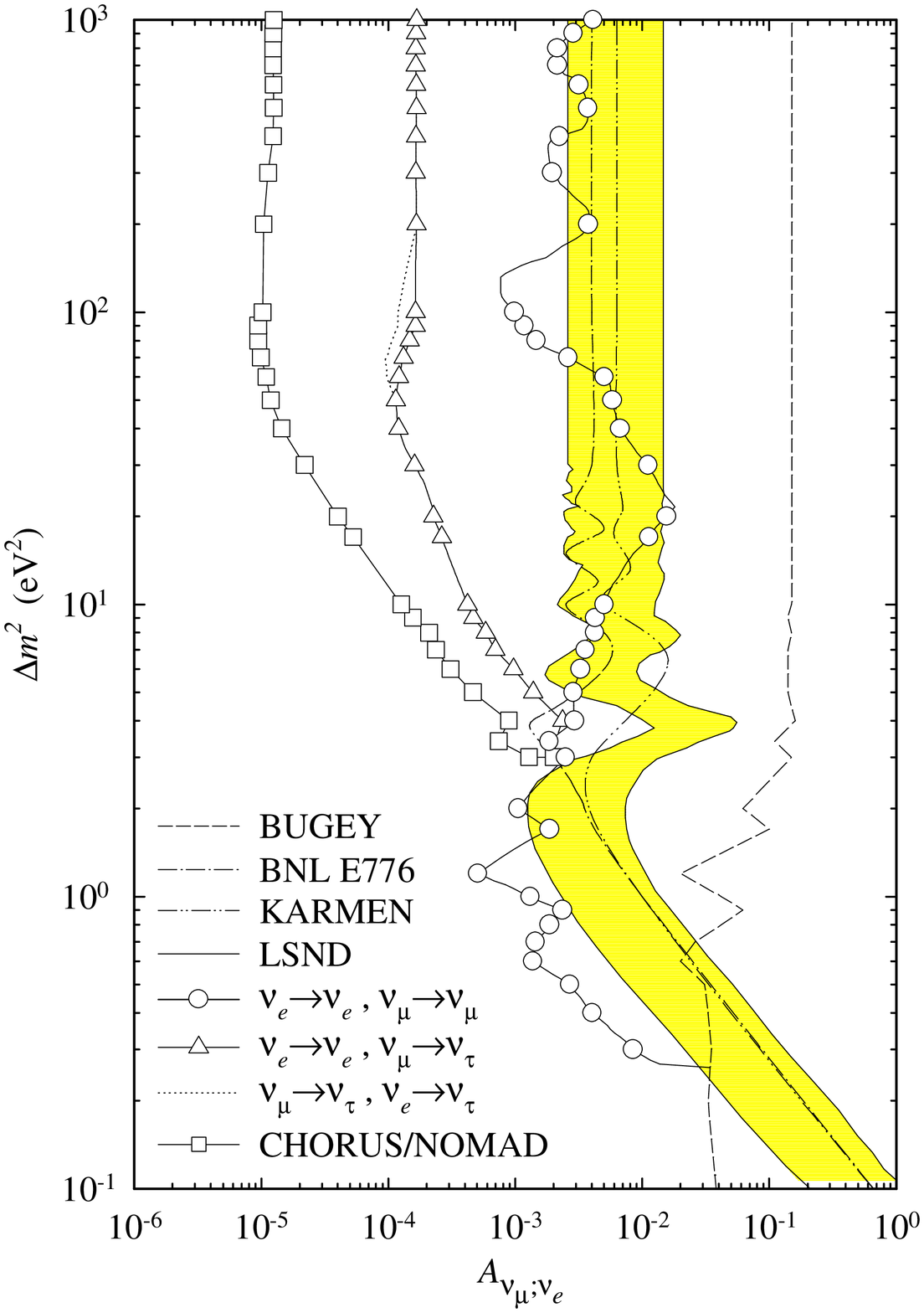,height=0.8\textheight}}
\end{center}
\protect\caption{Exclusion regions in the
$ A_{\nu_{\mu};\nu_{e}} $--$ \Delta m^2 $
plane
for small
$ \left| U_{e3} \right|^2 $
and
$ \left| U_{\mu3} \right|^2 $.}
\label{fig1}
\end{figure}

I will discuss now neutrino oscillation in these two regions.
In the region \ref{R1}
$ \nu_\mu \leftrightarrows \nu_e $
oscillations are suppressed.
In fact, from
Eq.(\ref{38}) we have
\begin{equation} 
A_{\nu_\mu ; \nu_e} \leq 4 \, a^0_\mu \, a^0_e
\label{49a}
\end{equation}
In the Fig.\ref{fig1}
this limit is presented by the curve passing through
circles.
We have also plotted in Fig.\ref{fig1}
the LSND allowed region
(the shadowed region
between two solid lines)
and the exclusion plots obtained in the KARMEN \cite{KARMEN} and
BNL E776 \cite{BNLE776} experiments.

Limits on the amplitude of
$ \nu_\mu \leftrightarrows \nu_e $
oscillations
more stringent than those given by Eq.(\ref{49a})
can be obtained if the exclusion
plot found in the FNAL E531 experiment \cite{FNALE531}
on the search for
$\nu_\mu \leftrightarrows \nu_\tau $
oscillations is taken into account.
In fact, in
linear approximation in the
small quantities $a^0_e$ and $a^0_\mu$ we have
\begin{equation}
A_{\nu_\mu ; \nu_\tau} \simeq 4 \, |U_{\mu3} |^2
\end{equation}
Thus, for the amplitude of
$ \nu_\mu \leftrightarrows \nu_e $
oscillations we have the following upper bound:
\begin{equation}
A_{\nu_\mu ; \nu_e} \leq A^0_{\nu_\mu ; \nu_\tau} \, a^0_e
\label{48}
\end{equation}
where
$ A_{\nu_\mu ; \nu_\tau} \leq A^0_{\nu_\mu ; \nu_\tau} $
and
$ A^0_{\nu_\mu ; \nu_\tau}$
can be obtained
(at any fixed value of 
$ \Delta m^2$)
from the exclusion plot found in the FNAL E531 experiment.
The upper
bound (\ref{48}) is presented in Fig.1
by the curve passing through triangles.
As
it is seen from Fig.1,
the LSND-allowed region is not compatible with the data of
all the other experiments
on the search for neutrino oscillations
(if there is a neutrino
mass hierarchy and the parameters $ |U_{e3}|$ and $ |U_{\mu3}| $
are both small).

In the linear approximation in
$a^0_e$ and $a^0_\mu$,
for the
amplitude of
$ \nu_e \leftrightarrows \nu_\tau $
oscillations we have
\begin{equation}
A_{\nu_e ; \nu_\tau} \simeq 4 \, |U_{e3}|^2
\label{49}
\end{equation}
From the relations (\ref{38}) and (\ref{49}),
it follows that,
in the region of
$\Delta m^2$ considered here,
the amplitudes of
$ \nu_\mu \leftrightarrows \nu_\tau$
and
$ \nu_e \leftrightarrows \nu_\tau $
oscillations must be small:
\begin{eqnarray}
&&
A_{\nu_\mu ; \nu_\tau} \leq 4 \, a^0_\mu
\\
&&
A_{\nu_e ; \nu_\tau} \leq 4 \, a^0_e
\end{eqnarray}
There are no other constrains on the amplitudes 
$A_{\nu_{\mu} ; \nu_{\tau}}$
and
$ A_{\nu_{e} ; \nu_{\tau}}$ from the
result of disappearance
experiments.
As it is well known,
the search for
$ \nu_\mu \leftrightarrows \nu_\tau $
(and also
$ \nu_e \leftrightarrows \nu_\tau $)
oscillations is
going on in CERN (CHORUS \cite{CHORUS} and NOMAD \cite{NOMAD}).
In the region of $ \Delta m^2 \geq 12 \, \mbox{eV}^2 $
these experiments will be sensitive to
$ A_{\nu_\mu ; \nu_\tau} \leq 2 \times 10^{-4} $,
which is much less than $ 4 \, a^0_\mu $.    

In the region \ref{R2},
due to unitarity of the mixing matrix,
\begin{equation}
|U_{e3}|^2 \leq 1- |U_{\mu3}|^2 \leq a^0_{\mu}
\label{51}
\end{equation}
Thus,
\begin{equation} 
|U_{e3}|^2 \leq 4 \, \mbox{Min}(a^0_e,a^0_\mu)
\label{52}
\end{equation}
For the amplitude of $ \nu_\mu \leftrightarrows \nu_e $ oscillations,
from the results of disappearance experiments
we have the following upper
bound
\begin{equation}
A_{\nu_\mu ; \nu_e} \leq
4 \, |U_{e3}| \leq 4 \, \mbox{Min}(a^0_e,a^0_\mu)
\end{equation}
This bound
(for $\Delta m^2 > 3 \times 10^{-1} \, \mbox{eV}^2 $)
is less
stringent than the bounds that were found in
the KARMEN and BNL E776 experiments.
For the amplitude of
$ \nu_\mu \leftrightarrows \nu_\tau $
oscillations,
from
the results of the disappearance experiments the same upper bound
as in
the case of region \ref{R1} can be obtained:
\begin{equation}
A_{\nu_\mu ; \nu_\tau} \leq 4 |U_{\tau 3}|^2 \leq 4 a^0_\mu
\end{equation}
If the parameters $|U_{e3}|^2$ and $|U_{\mu3}|^2$
are in the region \ref{R2},
the oscillations $ \nu_e \leftrightarrows \nu_\tau $
are strongly
suppressed.
In fact, from Eqs.(\ref{38}), (\ref{51}) and (\ref{52})
we have
\begin{equation}
A_{\nu_e ; \nu_\tau} \leq 4 \, \mbox{Min}(a^0_e ,a^0_\mu )
\ a^0_\mu
\end{equation}
This bound is much stronger than the bound that was obtained in
FNAL E531 experiment \cite{FNALE531}.

Thus, if the LSND result is confirmed by future experiments,
the parameters
$|U_{e3}|^2$ and $|U_{\mu3}|^2$
cannot be
both small
(assuming that there is a hierarchy of neutrino masses),
i.e. there is no hierarchy of couplings in the lepton sector.
The only
possible solution that is compatible with the LSND and all other neutrino
oscillation experiments is the solution with small
$ |U_{e3}|^2$ and large 
$|U_{\mu3}|^2$.
This "unnatural" neutrino mixing would mean that $\nu_\mu$ is
the "heaviest" neutrino.

\section{Solar neutrinos}

In conclusion we will consider briefly
the solar neutrino problem.
Probably at the moment the strongest indications in favour of neutrino
mixing come from solar neutrino experiments.
The reactions of the thermonuclear
$p-p$ cycle,
which are the main sources of solar $\nu_e$'s,
are listed in the
Table 2.

\begin{table}[b]
\label{T02}
\protect\caption{Main sources of solar $\nu_e'$s.}
\begin{center}
\begin{tabular}{|l|c|c|} 
\hline
Reaction &
Neutrino energies $E_\nu$ (MeV) &
Expected fluxes 
($ \mbox{cm}^{-2} \mbox{sec}^{-1}$)
\\ \hline
$ p \, p \rightarrow d \, {e^{+}} \, \nu_e $  &
$ \leq{0.42}$ &$  6.0 \times {10^{10}} $ \\
$ e^{-} \, {^{7}Be} \rightarrow {^{7}Li} \, \nu_{e} $ &
0.86 & $ 4.9 \times{10^9} $ \\
$ ^{8}B \rightarrow {^{8}Be} \, e^{+} \, \nu_e $ &
$ \leq{14}$ & $ 5.7 \times {10^6} $\\
\hline
\end{tabular}
\end{center}
\end{table}

In the last column of the 
Table 2 the fluxes calculated on 
the basis of the Standard Solar Model (SSM)
\cite{BSSM}
are
presented.
The fluxes of $^{8}B$ and other neutrinos depend strongly on
the cross
sections of nuclear reactions,
on the temperature of the sun and other input
parameters.
There is, however, a general constraint on
the solar neutrino fluxes.
The solar energy is produced in the transition
\begin{equation}
2 e^- + 4 p \rightarrow ^{4} He + 2 \nu_e
\end{equation}
Thus, the production of luminous energy of the sun is accompanied by the
emission of neutrinos.
If the sun is in a stable state,
we have the following
relation
\begin{equation}
\frac{1}{2} \, Q
\sum_{i=pp,\ldots}
\left( 1 - 2 \frac{\overline{E}_i}{Q} \right)
\phi_i =
L_\odot
\label{57}
\end{equation}
Here $ Q=4m_p + 2m_e - m_{^{4}He} \simeq 26.7 \, \mbox{MeV}$,
$L_{\odot}$ is the
luminosity of the sun,
$R$ is the distance between the sun and the earth,
$\phi_i$ is the total flux of neutrinos from the source 
$i$ ($i=pp,^{7}Be,\ldots$) and $\overline{E}_i$
is the average energy of the neutrinos from the source
$i$.
In the Table 3
the results of the four solar neutrino experiments
(Homestake \cite{Homestake},
Kamiokande \cite{Kamiokande},
GALLEX \cite{GALLEX}
and SAGE \cite{SAGE})
are
presented.

\begin{table}[b]
\label{T03}
\protect\caption{Solar neutrino data.}
\begin{center}
\begin{tabular}{|l|c|c|}
\hline
Experiment &
Counting rate (SNU) &
Predicted rate (SNU)
\\
\hline
Homestake & & \\
$ ( \nu_{e} {^{37} Cl} \rightarrow {^{37}Ar} e^{-} )$ &
$ 2.56 {}^{+0.17}_{-0.13} \pm 0.14 $ 
& $ 9.3 \pm 1.4 $
\\
GALLEX & & \\
$( \nu_{e} {^{71}Ga} \rightarrow {^{71}Ge} e^{-} )$ &
$ 77 \pm {8.5} \pm {5} $ &
$ 131.5 \pm {6} $
\\
SAGE & & \\
$( \nu_{e} {^{71}Ga} \rightarrow {^{71}Ge} e^{-} )$ &
$ 72 {}^{+12}_{-10} {}^{+5}_{-7} $ &
$ 131.5 \pm {6} $
\\
KAMIOKANDE & & \\
$ ( \nu \, e \rightarrow \nu \, e ) $ &
$ \frac{\mathrm{data}}{\mathrm{SSM}}
=
0.51 \pm {0.04} \pm {0.06} $ &
\\
\hline
\end{tabular}
\end{center}
\end{table}

As it is seen from the Table 3,
the observed event rates in all
experiments are significantly less than
the SSM expected rates.
A few
comments on this problem.

If we assume that
$ P_{\nu_e \rightarrow \nu_e} = 1 $,
from the luminosity
constraint (\ref{57})
for the neutrino counting rate in
the GALLEX and SAGE experiments
we find the following lower bound:
\begin{equation}
Q_{Ga}
=
\sum_{i} \overline{\sigma}_i \, \Phi _{i}
\geq
\overline{\sigma}_{pp}
\sum _{i}
\Phi _i \simeq 80 \, \mbox{SNU}
\label{58}
\end{equation}
where $\overline{\sigma}_i$
is the average value of the cross section of
the process
$ \nu_{e} + {^{71}Ga} \rightarrow e^{-} + {^{71}Ge} $.
This lower bound
does not contradict the rate measured in
the GALLEX and SAGE experiments.

Let us consider the data of
\emph{different} solar neutrino experiments.
We will assume only that $ P_{\nu_e \rightarrow \nu_e} =1 $.
In the
Kamiokande experiment,
due to the high energy threshold of the detected electrons
($\simeq 7 \, \mbox{MeV}$),
only $^{8}B$ neutrinos are detected.
From the data of this
experiment,
for the total flux of $^{8}B$ neutrinos it was obtained
\cite{Bahcall94}
\begin{equation}
\Phi^{^{8}B} = (2.89 \pm 0.21 \pm 0.35 )
\times 10^6
\, \mbox{cm}^{-2} \, \mbox{s}^{-1}
\label{59}
\end{equation}

The main contribution
to the event rate in the Homestake chlorine experiment
comes from $^{8}B$ and $^7Be$ neutrinos.
Using the value (\ref{59}) of the
$^{8}B$ flux for the contribution of
the $^{7}Be$ neutrinos
to the event rate of
the chlorine experiment,
we have the following upper bound:
\begin{equation}
Q_{Cl}(^{7}Be) \leq 0.46 \, \mbox{SNU}
\label{60}
\end{equation}
This upper bound is not compatible with the predictions of
the existing
Standard Solar Models:
\begin{equation}
Q_{Cl}^{\mathrm{SSM}}(^{7}Be)= 1.1 \pm 0.1 \, \mbox{SNU}
\label{61}
\end{equation}

The main contribution to the event rate of
the gallium experiments comes
from the
$pp$, $^{7}Be$ and $^{8}B$ neutrinos.
Using the luminosity constraint (\ref{57})
and the value (\ref{59}) of the $^{8}B$ neutrino flux,
for the contribution 
of $^{7}Be$ neutrinos to
the event rate in
the GALLEX experiment we have the following upper bound:
\begin{equation}
Q_{Ga}(^{7}Be) \leq 19 \, \mbox{SNU}
\end{equation}
This value is significantly smaller
than the values predicted by 
the existing Standard Solar Models:
\begin{equation}
Q_{Ga}^{\mathrm{SSM}}(^7Be) = 34 \pm 4 \, \mbox{SNU}
\end{equation}

Neutrino mixing is the most plausible explanation of the solar
neutrino data.
In fact, all the existing data can be explained if neutrino
mixing enhanced by the MSW matter effects
\cite{MSW}
is assumed.
For the mixing
parameters the following values were obtained
(see Ref.\cite{SOLMSW})
\begin{eqnarray}
&&
\Delta m^2 \simeq 5\times 10^{-6} \mbox{eV}^2,
\qquad
\sin ^2{2\theta}\simeq 8\times 10 ^{-3}
\\
\mbox{or}
&&
\nonumber
\\ 
&&
\Delta m^2 \simeq 5\times 10^{-5} \mbox{eV}^2,
\qquad
\sin ^2{2\theta} \simeq 0.8
\end{eqnarray}
These values were obtained under the assumption that solar
neutrino fluxes are given by the BP \cite{BSSM} SSM model.  
 
The solar neutrino experiments at the moment give
us the most
compelling indications in favour of neutrino mixing.
However, new experimental
data are needed to obtain
an evidence for neutrino masses and mixing.

In conclusion,
I will mention some model independent possibilities to
obtain information about neutrino mixing from future solar neutrino
experiments
\cite{BG93,BG94}.
In the Super-Kamiokande experiment,
scheduled for 1996 \cite{SK}
solar neutrinos will be detected through the observation
(about 40 events/day)
of the $ES$ process
$ \nu \, e \rightarrow \nu \, e $.
In the SNO
experiment,
scheduled for 1997 \cite{SNO},
solar neutrinos will be detected 
through the observation of the $CC$ process
$ \nu_e \, d \rightarrow e^- \, p \, p $,
the $NC$
process
$ \nu \, d \rightarrow \nu \, n \, p $
and also the $ES$ process
$ \nu \, e \rightarrow \nu \, e $.
In both experiments,
due to the high energy threshold,
only $^{8}B$
neutrinos will be detected.
The flux of initial $^{8}B$ neutrinos is given by
\begin{equation}
\Phi^0_{\nu_e}(E)
=
X(E) \, \Phi^0_{B}
\end{equation}
where $\Phi^0_{B}$ is the unknown total flux and $X(E)$ is
the known
function of neutrino energy $E$
(determined mainly by the phase space factor of the
decay
$ {^{8}B} \rightarrow {^{8}Be} \, e^+ \, \nu_e $).
The detection of solar
neutrinos with the observation of
the $CC$, $NC$ and $ES$ processes will allow to
analyze the content of the beam of solar neutrinos on the earth,
to
determine the total flux $\Phi_B$ and
the probability of solar neutrinos to
survive $P_{\nu_e \rightarrow \nu_e}(E)$.
In particular, it will
be possible to check
\cite{BG94}
in a model independent way
if there are transitions of
solar $\nu_e'$s into sterile states.
Indeed, the spectrum of the
recoil electrons in the process
$ \nu \, e \rightarrow \nu \, e $
can be presented
in the form
\begin{equation}
n^{ES}(T)
=
\int_{E_{\mathrm{min}}(T)}
\left[
\left(\frac{d\sigma}{dT}\right)_{\nu_{e}e}
-
\left(\frac{d\sigma}{dT}\right)_{\nu_{\mu}e}
\right]
\Phi_{\nu_e}(E)
\, dE
+
\Sigma^{ES}(T)
\end{equation}
where $T$ is electron kinetic energy,
$(d\sigma/dT)_{\nu_{l}e}$
is the differential cross section of the process
$ \nu_l \, e \rightarrow \nu_l \, e$,
$ E_{\mathrm{min}} = T ( 1 + \sqrt{ 1 + 2m_e / T } ) / 2 $,
$\Phi_{\nu_e}(E)$ is the spectrum of $\nu_e$'s on the earth,
that will be
measured in the SNO experiment, and
\begin{eqnarray}
\Sigma^{ES}(T)
&=&
\int_{E_{\mathrm{min}}(T)}
\left(\frac{d\sigma}{dT}\right)_{\nu_{\mu}e}
\sum_{l=e,\mu,\tau}
P_{\nu_{l} \rightarrow \nu_{l}}
\, X(E) \, dE \ \Phi _B
\nonumber\\
&=&
\left\langle
\sum_{l=e,\mu,\tau} P_{\nu_l \rightarrow \nu_l}
\right\rangle_{T} 
X_{\nu_{\mu}e}(T) \, \Phi_B
\end{eqnarray}
where
\begin{equation}
X_{\nu_{\mu}e}(T)
=
\int_{E_{\mathrm{min}}(T)}
\left(\frac{d\sigma}{dT}\right)_{\nu_{\mu}e}
\, X(E) \, dE
\end{equation}
is a known function of $T$.
The function $\Sigma^{ES}(T)$ can be
determined from experiment by the measurement
of $n^{ES}(T)$ in the $ES$ process
and $\Phi_{\nu_e}(E)$ in the $CC$ process.
If the function 
$ \Sigma^{ES}(T) / X_{\nu_{\mu}e}(T) $
depends on $T$, it means
that the total probability of the transition of $\nu_e$'s into active 
states
$\sum_{l=e,\mu,\tau} P_{\nu_{e} \rightarrow \nu_{l}}$
is 
less than one, i.e.
that
the solar $\nu_e$'s are transformed into sterile states.

There exist also the atmospheric neutrino anomaly observed by
the Kamiokande \cite{Kamiokande-atmospheric},
IMB \cite{IMB} and Soudan 2 \cite{Soudan} experiments.
This anomaly can
be explained with
$ \nu_{\mu} \leftrightarrows \nu_e $
or
$ \nu_{\mu} \rightarrow \nu_\tau $
neutrino oscillations with
$ \sin ^2 2\theta \simeq 1 $
and
$ \Delta m^2 \simeq 10^{-2} \, \mbox{eV}^2 $.
Several long-baseline terrestrial
neutrino experiments now in preparation will be sensitive to neutrino
oscillations with
$ \Delta m^2 \simeq (10^{-2}-10^{-3}) \, \mbox{eV}^2 $
and 
$ \sin^2 2\theta \gtrsim 0.1 $
\cite{MINOS,ICARUS,KEK}.

In conclusion, the problem of neutrino masses and mixing raised
many years ago by B. Pontecorvo
is the key problem of today's neutrino physics.
We have at the moment different indications
in favour of non-zero neutrino
masses and mixing angles.
Future experiments could be decisive for this very
important problem of physics and astrophysics. 

I would like to express my gratitude to Carlo Giunti for fruitful
collaboration and very useful discussions.

\end{document}